\documentclass[journal]{IEEEtran}

\usepackage{amsmath,amssymb}
\usepackage{booktabs}
\usepackage{graphicx}
\usepackage{url}
\usepackage[hidelinks,breaklinks]{hyperref}

\begin{document}

\title{Active Inference for Slow and Fast Interaction\\
Loops in Holographic-Type Communication}

\author{Kamran~Fotovat,
        Kian~Fotovat,
        Nika~Hosseini,
        and~Martin~Maier
\thanks{Kamran Fotovat is with Iran Univ. of Science and Technology,
Tehran, Iran. Kian Fotovat is with Univ. of Tehran, Tehran, Iran. Nika
Hosseini and Martin Maier are with INRS, Montr\'eal, QC, Canada.}}

\markboth{Preprint}%
{Fotovat \MakeLowercase{\textit{et al.}}: Active Inference for Slow and Fast Interaction Loops in HTC}

\maketitle

\begin{abstract}
Holographic-type communication requires a new hologram when the viewer
moves beyond the display's angular viewing zone, whereas image-to-3D
reconstruction takes tens of seconds per asset. We separate these
operations into a slow loop that reconstructs a persistent mesh and a
fast loop that renders each viewpoint from the stored mesh. The fast loop
takes $0.117$~s independently of camera motion, compared with
$13.8$--$14.8$~s for a full
reconstruction. An active-inference controller determines when to begin
reconstruction from learned user-state transitions, whether to perform a
full or regional update, and how much computation to allocate based on a
monocular-depth detail measure. Regional updates preserve geometry
outside the edited area. In a modeled $100$-event session, separating the
loops reduces mean latency per event from $14.8$ to $4.4$~s
($3.4$-fold); predictive reconstruction reduces it further to
$1.7$--$2.9$~s, depending on user routine strength.
\end{abstract}


\section{Introduction}
\IEEEPARstart{H}{olographic}-type communication (HTC) promises
interaction with remote three-dimensional content at the fidelity of
physical presence. What makes it a communication and control problem
rather than a purely optical one is timing. A phase-only spatial light
modulator presents one hologram at a time, valid across an angular zone
only a few degrees wide, so movement beyond that zone demands a new one
within an interactive budget. Multi-view diffusion \cite{mvdream} and
image-to-3D generators \cite{trellis}, \cite{hunyuan3d2} can require tens
of seconds per asset and therefore cannot be used for every viewpoint
update.

In interactive HTC, however, a change of viewpoint does not require
reconstructing the underlying asset when its content is unchanged.
Content changes require more expensive work, whose cost depends on the
extent of the edit and the geometric detail of the object.
Anticipating a likely change also creates an opportunity to begin
reconstruction during an idle period. We therefore treat the timing,
scope, and computational budget of reconstruction as decisions.

Arranging the pipeline so that viewpoint changes do not invoke the
generator leaves content changes as the principal source of latency. This
residual cost is not constant: it depends on how much fine structure the
object carries, how much of it actually changed, and whether
reconstruction could have begun before the change was requested. Treating
it as fixed overspends on simple objects, under-resolves detailed ones,
and pays in full for an edit confined to a small region.

Our contributions are twofold. (i) We develop a slow- and fast-loop
architecture in which a persistent mesh is reconstructed when the content
changes, while new viewpoints are rendered from that mesh without
repeating image-to-3D reconstruction. (ii) We develop an
active-inference controller that anticipates likely content changes,
chooses between full and regional reconstruction, and adjusts the
computational effort to the content's geometric detail. These decisions
are brought together within an expected-free-energy formulation.

The remainder is organized as follows. Section~\ref{sec:model} presents
the system model and the two interaction loops; Section~\ref{sec:ctrl}
develops the active-inference controller and its three decisions;
Section~\ref{sec:results} reports the results; Section~\ref{sec:future}
sets out the work that follows; and Section~\ref{sec:conc} concludes.

\section{System Model: Slow and Fast Interaction Loops}
\label{sec:model}

\begin{figure*}[!t]
\centering
\includegraphics[width=\textwidth]{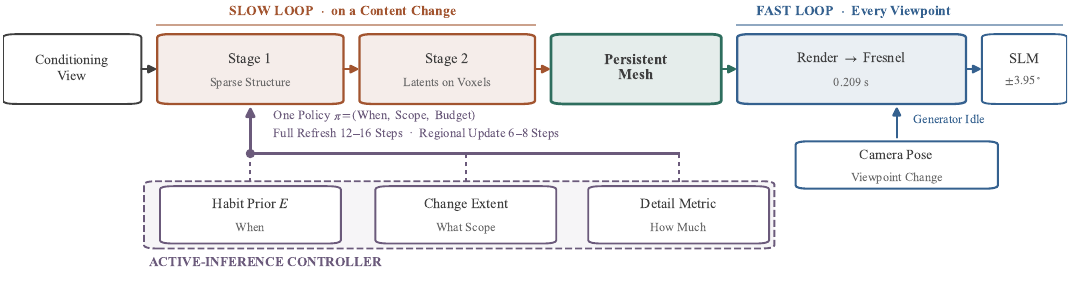}
\caption{The slow- and fast-loop pipeline. The generator runs only on a
content change; every viewpoint is served from the persistent mesh by
view rendering and Fresnel propagation. The controller allocates the slow
loop according to timing, scope, and computational budget.}
\label{fig:pipeline}
\end{figure*}

We split the pipeline where cost and required rate diverge
(Fig.~\ref{fig:pipeline}). The slow loop reconstructs an explicit mesh
from one conditioning view with TRELLIS \cite{trellis}, a
structured-latent generator. The fast loop renders intensity and depth
from that mesh at a given pose, accumulates $16$ depth slices, and
propagates each by Fresnel transform \cite{maimone}, \cite{shi_nature}
into an $8$-bit
phase-only map for a $3840\times2160$ modulator ($p=3.74~\mu$m,
$\lambda=515$~nm, $50$--$150$~mm depth range).

A phase-only modulator displays one hologram at a time, and diffraction
off a pitch-$p$ aperture limits the reconstructed field to a half-angle
\begin{equation}
\theta_{\max}=\arcsin\left(\frac{\lambda}{2p}\right)\approx3.95^\circ ,
\label{eq:zone}
\end{equation}
a total viewing zone of approximately $7.9^\circ$, or a $41$~mm eyebox at
$300$~mm. Within it one hologram gives continuous parallax and correct
accommodation with no recomputation; outside it the viewer sees nothing.
The per-viewpoint update is imposed by the display, not chosen by us.

What it does not require is a new 3D asset. The mesh being
viewpoint-invariant, a viewpoint change reduces to rendering intensity
and depth from the new pose and re-propagating the field. Both are
deterministic, closed-form and free of generative sampling, so reuse is
exact: one mesh held across $720$ consecutive frames produced no
measurable degradation, and no error accumulates however long it is kept.

\begin{figure}[!t]
\centering
\includegraphics[width=\columnwidth]{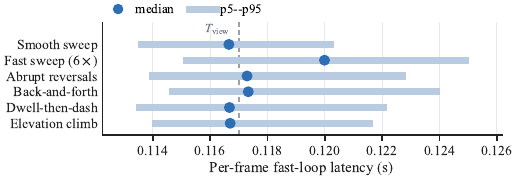}
\caption{Fast-loop latency over six camera trajectories, $119$ frames
each after a warm-up frame. Bands are p$5$--p$95$ within a trajectory and
overlap almost completely: the $2.9\%$ spread between medians is no larger
than the $2.0$--$2.8\%$ frame-to-frame variation inside each, so the cost
does not depend on how the camera moves. Dashed: the $T_{\mathrm{view}}$
of Table~\ref{tab:cost}.}
\label{fig:motion}
\end{figure}

That cost is constant, not a distribution: over the six camera
trajectories of Fig.~\ref{fig:motion} the medians span $2.9\%$, within
the $2.0$--$2.8\%$ frame-to-frame variation, so no motion dependence is
resolvable. Fresnel propagation dominates, and its cost is set by the
modulator, not the scene ($r\le0.19$ against both slice occupancy and
camera speed).

\begin{table}[!t]
\caption{Cost of Each Operation on an RTX~4070~SUPER (12~GB) With
Sequential CPU Offload. Generator Rows Are Medians of Five Interleaved
Repetitions, Excluding Weight Loading. Compute Sums the Per-Stage GPU
Timings; the Balance of Wall Time Is Host--Device Transfer. Peak Memory
$5.99$--$6.00$~GB.}
\label{tab:cost}
\centering
\small
\begin{tabular}{@{}llrrc@{}}
\toprule
Operation & Steps & Compute (s) & Wall (s) & Range (s)\\
\midrule
Viewpoint update & --- & \multicolumn{2}{c}{\textbf{0.117}$^{\dagger}$}
& ---\\
\midrule
Regional update & \phantom{1}6 & 2.03 & 10.12 & 9.80--11.05\\
Regional update & \phantom{1}8 & 2.11 & 10.20 & 9.79--11.06\\
Full refresh    & 12 & 5.21 & 13.78 & 13.72--14.28\\
Full refresh    & 16 & 6.38 & 14.76 & 14.31--15.12\\
\bottomrule
\end{tabular}

\vspace{2pt}
{\footnotesize $^{\dagger}$Median over the $720$ frames of
Fig.~\ref{fig:motion}. The fast loop holds no generator weights and so
incurs no offload transfer, which is why no separate compute and wall
figure is reported for it.}
\end{table}

Two ratios in Table~\ref{tab:cost} organize the design. A viewpoint
change is $118$ times cheaper than the least expensive full refresh, which
is why removing the generator from the interaction loop---rather than
merely accelerating it---makes interaction possible at all. And a
regional update is approximately $1.4$ times cheaper in wall time than a
full refresh at the corresponding budget, and $2.5$ to $3.0$ times
cheaper in computation. They differ because wall time is not proportional
to step count: every call pays a near-constant $8.1$--$8.6$~s of weight
transfer whatever its budget, above a compute term that does scale
($2.03$~s at $6$ steps, $6.38$~s at $16$). A $16$-step full refresh thus
costs $1.5$ times a $6$-step regional update in wall time but $3.1$ times
the computation, which is what makes the scope decision worth taking.

The backend is chosen for controllability, not raw speed. Faster
generators exist, including step-distilled models, but expose no useful
knobs: a distilled denoiser is trained for a fixed step count, and a
generator whose latent is an unordered set carries no spatial index, so
``regenerate only this region'' cannot be expressed. TRELLIS denoises a
dense $16^3$ grid into an occupancy field, then latents on those active
voxels. A region of space is thus a subset of tensor entries, and each
stage's step budget a free parameter. What reuse cannot remove is the
cost of a content change, signaled only as it occurs; that residual
uncertainty is temporal rather than optical, and it is where we apply
active inference.

\section{The Active-Inference Controller}
\label{sec:ctrl}
We cast this as active inference rather than reinforcement learning for
three reasons. There is no training budget: the controller faces a new
user from the first interaction, and every exploratory action costs that
user real waiting time. The quantities that should drive the decision are
already in physical units---an expected latency, a residual geometric
error, transition counts---whereas a reward would compress latency
against fidelity into one engineered scalar, precisely the tradeoff we
want explicit. Learning the user reduces to a Dirichlet update
\eqref{eq:bupd}: it adapts within a few dozen events, is inspectable, and
has one interpretable rate.

\begin{figure}[!t]
\centering
\includegraphics[width=\columnwidth]{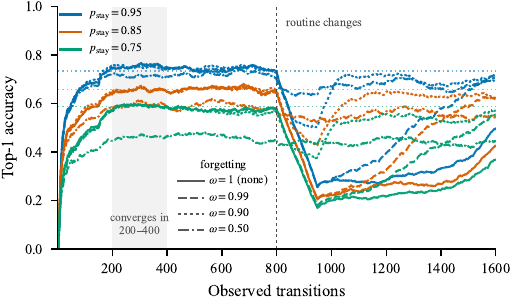}
\caption{Habit learning, $12$ states, mean of $10$ seeds, three routine
strengths (color) crossed with four forgetting rates (line style); dotted
horizontals are the corresponding Bayes-optimal ceilings. When the
routine changes at transition $800$, undecayed counts resist revision
while faster forgetting recovers sooner. The rates bracket the adopted
$\omega=0.90$: by $\omega=0.50$ the pre-change plateau no longer reaches
the ceiling.}
\label{fig:learning}
\end{figure}

\subsection{Detail Demand: Measured Before Reconstruction}
\label{sec:detail}
The step budget must be set from the conditioning view, before any
reconstruction exists to inspect. Measuring high-frequency content in the
image itself fails because appearance and geometry are conflated: a
smooth apple with streaked coloring scores $0.084$ against $0.084$ for a
dragon bust covered in scales, inverting the ordering a viewer would
give. We therefore measure on geometry. A monocular depth map is
predicted from the conditioning view by Depth Anything V2
\cite{depthanything}, converted to surface normals, band-passed, and
scored by the fraction of the surface carrying fine normal variation.
Depth carries no albedo, so painted texture drops out, and coverage
rather than magnitude is scored so a few silhouette edges cannot stand in
for detail spread over a surface. Across eight objects it orders them as
a viewer would---$0.133$ for a
carved bust and $0.116$ for the dragon, against $0.093$ for the same bust
unornamented and $0.060$ for the apple---and the controller acts on that
separation: a detailed object is reconstructed at $16$ steps, a simple
one at $12$, regional updates at $8$ or $6$. It costs about $0.05$~s
against tens of seconds for the reconstruction it informs.

\subsection{Scope: Regional Update}
\label{sec:scope}
Because stage~1 denoises a dense grid indexed by space, a region of the
object is a subset of tensor entries and can be regenerated alone.
Following \cite{repaint}, we re-noise the cached latents to a chosen
depth and, at every denoising step, overwrite the region to be kept with
the original latent re-noised to the current timestep. New geometry
appears only inside the mask while the generator still attends to the
surroundings, which is what makes the new surface meet the old.

We quantify geometric preservation with the Chamfer distance between
point samples from the original and updated mesh surfaces. On a paired
edit it was $3.9\times10^{-4}$ outside the edited region and
$1.0\times10^{-2}$ inside it, a $26$-fold difference. Two
consequences follow. An asset generated at a high budget keeps that
quality wherever it was not edited, amortizing the reconstruction over
later updates; and starting from a converged latent rather than noise
needs a shorter schedule, $6$--$8$ steps against $12$--$16$. The region
is localized by the test that detects the content change: its difference
map is nonzero exactly where the object no longer matches the mesh.

\subsection{Timing: Anticipating the Refresh}
\label{sec:timing}
A content change is signaled only when it occurs, so reconstruction
begins already late. It need not: if the next state can be predicted, the
work can start during the current idle period. Let $s_t$ index the
object's state and $B_{ji}=P(s_{t+1}{=}j\mid s_t{=}i)$ denote the
transition matrix. Under active inference \cite{friston}, \cite{parr},
\cite{namjoshi}
a policy $\pi$ follows
\begin{equation}
Q(\pi)=\sigma\bigl(\ln E-F^{[\pi]}-\gamma\,G^{[\pi]}\bigr),
\label{eq:pi}
\end{equation}
where $E$ is a prior over policies---the habit term---$G^{[\pi]}$ the
expected free energy scoring a future that has not happened, $F^{[\pi]}$
the evidence for the policy actually enacted, and $\gamma$ a precision on
$G$. We learn $E$ by Bayesian updating of $B$ under a Dirichlet prior
with concentration counts $\mathbf{b}$ \cite{dadrichlet}, using the
standard update with learning rate $\kappa$ and forgetting rate $\omega$,
\begin{equation}
\mathbf{b}\leftarrow\omega\,\mathbf{b}
+\kappa\,\textstyle\sum_\tau \mathbf{s}^{(\tau)}\!\circ \mathbf{s}^{(\tau-1)},
\label{eq:bupd}
\end{equation}
with $\omega=0.90$, chosen by sweeping $\omega$ from $1$ to $0.5$: it
raises accuracy over the $200$ events after a change of routine by a
third for under $1\%$ of steady-state accuracy, while faster rates erode
the steady state. Indexing a speculative
policy by the state it prepares for makes the transition posterior
directly usable as a policy prior:
\begin{equation}
\hat{P}(j\mid i)=\frac{b_{ij}+\eta\,P_0(j)}{b_{i\cdot}+\eta+\alpha},
\qquad
\ln E_{\mathrm{spec}}(j)\;\propto\;\ln\hat{P}(j\mid i),
\label{eq:pred}
\end{equation}
where $P_0$ is the user's global marginal, giving backoff of strength
$\eta$ for rarely visited states, and $\alpha$ reserves mass for an
unseen successor, keeping the state space open.

Crucially, we predict the next action, not the state at a future time. A
time-indexed formulation forces a horizon of $13$--$24$ steps, at which
the Bayes-optimal top-1 accuracy of a routine process collapses to
approximately $0.19$; at a horizon of one event it admits approximately
$0.73$.

\subsection{One Objective Over the Decisions}
\label{sec:efe}
A policy is the triple (when, scope, budget), and the expected free
energy of \eqref{eq:pi} scores all three at once:
\begin{equation}
G^{[\pi]}=\underbrace{\mathbb{E}[L]}_{\text{risk}}
\;+\;\underbrace{\beta\,\varepsilon(n,d)}_{\text{ambiguity}} .
\label{eq:efe}
\end{equation}
Risk is the deviation from the preference for low latency and carries the
timing and scope decisions; ambiguity is the residual geometric error
$\varepsilon$ left by $n$ steps on an object of detail $d$, and carries
the budget decision. We include no epistemic term: active inference
admits one, but every exploratory action is paid for in a user's waiting
time, and Eq.~\eqref{eq:bupd} learns the model passively from events that
occur anyway.

Writing $p$ for the predictor's top-1 accuracy, $T$ for the cost of the
chosen operation, $D$ for the idle time remaining and $t_{\mathrm{swap}}$
for the cost of adopting a finished result, risk reduces to an expected
latency
\begin{equation}
\mathbb{E}[L]=p\,\mathbb{E}\!\left[\max(t_{\mathrm{swap}},\,T-D)\right]
+(1-p)\,T .
\label{eq:habitlat}
\end{equation}
Scope enters through $T$, timing through $D$. Two properties follow: this
is never worse than not speculating, since a miss costs exactly $T$, and
speculation need not finish, because partial progress on a correct guess
is retained. The budget enters differently---we do not have
$\varepsilon(n,d)$ calibrated, so the detail measure supplies $d$ and
hence a prior, used to select among the budgets of Table~\ref{tab:cost}
rather than to compute an optimum (Section~\ref{sec:future}).

\section{Results}
\label{sec:results}

\begin{figure}[!t]
\centering
\includegraphics[width=\columnwidth]{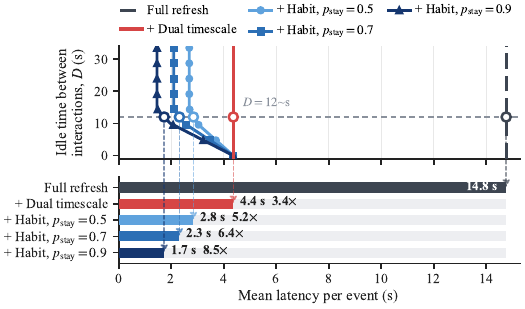}
\caption{Mean latency per event over a $100$-event session, from
\eqref{eq:habitlat} at the costs of Table~\ref{tab:cost}. Above: the sweep
over idle time $D$, latency on the horizontal axis so both panels share
it. Below: the readout at the $D=12$~s operating point, each bar joined to
its own crossing; the pale remainder of each bar is the latency that
configuration removed. Neither baseline speculates, so both are flat in
$D$ and appear above as vertical lines. The session is $65\%$ viewpoint,
$18\%$
regional and $17\%$ full; $p_{\mathrm{stay}}$ enters only through the
predictor's measured accuracy, $0.39$, $0.53$ and $0.68$.}
\label{fig:session}
\end{figure}

Predictor results use synthetic $12$-state Markov processes with known
transition matrices, scored prequentially: each event is predicted before
it is learned from, so hit rate is swept, not fitted.

Eq.~\eqref{eq:pred} attains $95$--$100\%$ of the closed-form Bayes-optimal
one-step ceiling on first-order chains, converging within $200$--$400$
transitions (Fig.~\ref{fig:learning}). Because a first-order model scored
on a first-order process is partly self-fulfilling, we repeat the
comparison on families it does not match---second-order chains,
semi-Markov dwells, drift, cyclic workflows---against references estimated
in hindsight. It recovers $97$--$100\%$ of the first-order optimum on
every family and $208\%$ under drift, beating a stationary oracle because
forgetting tracks a process no fixed table represents.

Fig.~\ref{fig:session} composes the three decisions over $100$ events.
Serving every event by full refresh costs $14.8$~s. The slow- and
fast-loop decomposition alone brings that to $4.4$~s, a factor of $3.4$,
almost all of it from removing the generator from the viewpoint path.
Anticipation removes a further third to three-fifths of what
remains---$2.9$~s at a weakly routine user and $1.7$~s at a strongly
routine one, $8.5$ times the baseline---acting only on the slow loop,
since a $0.117$~s viewpoint update has nothing worth hiding.

That return depends on the user's pace, which the upper panel of
Fig.~\ref{fig:session} sweeps. At $D=0$ every habit curve meets the
dual-timescale line at $4.4$~s---with nothing to speculate in the
mechanism returns nothing, not a free saving---and by $D=20$~s they reach
$2.7$, $2.1$ and $1.5$~s, each flattening once $D$ exceeds the cost it
hides. The $12$~s operating point lies on the steep part of every curve.

The decisions compound: the gain runs through $D/T$, so a cheaper scope
also moves the operation within reach of an idle period. At $D=8$~s
anticipation returns $1.45\times$ on a full refresh, $1.73\times$ on a
regional update and $2.18\times$ at a routine user, where the two
effects multiply.

\section{Future Work}
\label{sec:future}
Several extensions follow from the architecture. Calibrating
reconstruction error $\varepsilon(n,d)$ against step count and detail
demand would let the budget decision in \eqref{eq:efe} solve for an
allocation directly rather than select among those of
Table~\ref{tab:cost}, exposing a continuous fidelity--latency trade-off.

Because a viewpoint update never invokes the generator, the two loops can
be scheduled against one another rather than alternated: running the slow
loop continuously behind the viewpoint stream would raise the fraction of
content changes already in progress when requested. A resident-model
deployment would further widen the separation between budgets, presently
$0.08$~s in wall time against a $3.1$-fold spread in computation. Faster
hardware compounds with the controller rather than displacing it, since
shrinking $T$ raises $D/T$. Finally, because the predictor learns
transitions over an abstract user state, the same mechanism extends
unchanged to multi-user sessions.

\section{Conclusion}
\label{sec:conc}
We reduced interaction latency in HTC by separating viewpoint rendering
from 3D reconstruction: a slow loop updates a persistent mesh on a
content change, and a fast loop serves every viewpoint from it in
$0.117$~s against $13.8$--$14.8$~s for a full reconstruction. An
active-inference controller sets the timing, scope and budget of each
reconstruction, so that the three decisions are taken against one
objective rather than tuned independently. Over a modeled $100$-event
session this cut mean latency from $14.8$ to $4.4$~s per event, and to
$1.7$--$2.9$~s once anticipation is included. The decomposition is what
makes the residual cost addressable: with the generator off the viewpoint
path, latency concentrates in content changes, which are predictable,
bounded in scope, and schedulable into idle periods. These gains are
established
component-wise and through the latency model of \eqref{eq:habitlat}, so
the ratios they report should be read as properties of the structure
rather than of any one deployment.

\end{document}